\documentclass[12pt]{article}
\newlength{\abstractwidth}
\pdfoutput=1

\usepackage{epsf}
\usepackage{color}
\usepackage{graphicx}

\renewcommand{\thefootnote}{\fnsymbol{footnote}}
\renewcommand{\thanks}[1]{\footnote{#1}}
\newcommand{\starttext}{
\setcounter{footnote}{0}
\renewcommand{\thefootnote}{\arabic{footnote}}}

\newcommand{\bea}{\begin{eqnarray}}
\newcommand{\eea}{\end{eqnarray}}
\newcommand{\ee}{\end{equation}}
\newcommand{\be}{\begin{equation}}

\newcommand{\sm}{\smallskip}

\def\cC{{\cal C}}

\def\bR{{\bf R}}

\def\l({\left(}
\def\r){\right)}

\def\a{\alpha}

\def\sh{{\rm sh}}

\def\Bh{\hat{B}}
\def\Th{\hat{T}}
\def\sh{\hat{s}}

\def\no{\nonumber}

\begin{document}
\starttext
\setcounter{footnote}{0}

\begin{flushright}
\today
\end{flushright}

\bigskip

\begin{center}

{{\Large \bf Holographic Metamagnetism, Quantum Criticality,} \\
 \medskip
{\Large \bf  and Crossover Behavior}  
\footnote{This work was supported in part by NSF grant PHY-07-57702.}}

\vskip .4in

{\large \bf Eric D'Hoker and  Per Kraus}

\vskip .2in

{ \sl Department of Physics and Astronomy }\\
{\sl University of California, Los Angeles, CA 90095, USA}\\
{\tt \small dhoker@physics.ucla.edu; pkraus@ucla.edu}

\end{center}

\vskip .2in

\begin{abstract}

\vskip 0.1in 

Using high-precision numerical analysis, we show that $3+1$ dimensional 
gauge theories holographically dual to $4+1$ dimensional 
Einstein-Maxwell-Chern-Simons theory undergo a quantum phase transition
in the presence of a finite charge density and magnetic field.   The quantum critical theory
has dynamical scaling exponent $z=3$, and is reached by tuning a relevant operator of
scaling dimension $2$.  For magnetic field $B$ above the critical value $B_c$, 
the system  behaves as a  Fermi liquid.   As the magnetic field 
approaches $B_c$ from the high field side, the specific heat coefficient diverges as $1/(B-B_c)$, 
and non-Fermi liquid behavior sets in.   
For $B<B_c$  the entropy density $s$ becomes 
non-vanishing at zero temperature, and scales according to 
$s \sim \sqrt{B_c - B}$. At  $B=B_c$, and for
small non-zero temperature $T$, a new scaling law sets in for which $s\sim T^{1/3}$. Throughout
a small region surrounding the quantum critical point, the ratio 
$s/T^{1/3}$ is given by a universal scaling 
function which depends only on the ratio $(B-B_c)/T^{2/3}$.

\sm
 
The quantum phase transition involves non-analytic behavior of the specific heat and magnetization
but no change of symmetry.  Above the critical field, our numerical results are consistent with those
predicted by the Hertz/Millis theory applied to metamagnetic quantum phase transitions,
which also describe non-analytic changes in magnetization without change of symmetry.  
Such transitions have been the subject of much experimental investigation recently, 
especially in the compound Sr$_3$Ru$_2$O$_7$, and we comment on the connections.

\end{abstract}

\newpage


\newpage

\section{Introduction and summary of results}
\setcounter{equation}{0}

The AdS/CFT correspondence provides a precise and powerful tool for the
study of thermodynamics, statistical mechanics, and transport properties in 
a variety of 4-dimensional gauge theories at finite temperature,  charge density, 
and magnetic field. In the large $N$ and large `t~Hooft coupling limits these 
gauge theories are holographically dual to certain black brane solutions
(with both electric and magnetic charges) to 5-dimensional Einstein-Maxwell theory.
This theory includes a Chern-Simons term, whose coupling $k$ captures the strength 
of the chiral anomaly in the Maxwell current. For the special value $k=2/\sqrt{3}$ 
(in our conventions) the bulk Einstein-Maxwell theory is a consistent supersymmetric 
truncation of Type~IIB or M-theory \cite{Buchel:2006gb,Gauntlett:2006ai,Gauntlett:2007ma}, 
while for $k$ not equal to this value, supersymmetry is lost.

\sm

The study of thermodynamics and transport properties in 4-dimensional Yang-Mills
theory is of direct interest to the physics of heavy ion collisions at RHIC (and soon at the LHC), where 
quarks and gluons are subject to high temperatures, strong magnetic fields, and 
large currents.  Of great interest as well is the possibility of studying novel phases of finite
density matter at low temperatures.  By varying the charge density $\rho$  and magnetic field $B$ one
can search for signals of  zero temperature quantum phase transitions \cite{sachdev}, as is done experimentally 
in, for example, the heavy fermion compounds and high temperature superconductors. 

\sm

The  thermodynamics of the relevant solutions to 5-dimensional Einstein-Maxwell
theory can be understood analytically in two limiting cases. 
In the absence of magnetic fields, the relevant  supergravity solution is the 
electrically charged (Reissner-Nordstrom) black brane in $AdS_5$. At high temperature,
its entropy density $s$ scales as $T^3$, as dictated by scale invariance in the UV, 
and in accord with weak coupling Yang-Mills theory. 
As $T$  tends to 0, the black brane tends to its extremal form with 
$AdS_2 \times \bR^3$ near-horizon geometry, and non-vanishing  entropy density.
On the one hand, this AdS$_2$ factor is needed in  ``semi-holographic" 
models of non-Fermi liquid behavior \cite{Liu:2009dm,Cubrovic:2009ye,Faulkner:2009wj,Faulkner:2010tq} (see also \cite{Rey:2008zz}).  
On the other hand the ground state entropy density
is rather exotic from the Fermi surface perspective. It is exotic also from the point of view
of the CFTs that arise in the AdS/CFT duality, which typically contain massless charged bosons
that would be expected to condense and lead to a unique ground state. 

\sm

The other limiting case consists of vanishing electric charge density and finite magnetic field $B$. 
At high temperature the entropy density $s$ again scales as $ T^3$, 
but at low temperature the analysis of  \cite{D'Hoker:2009mm} showed that $s\sim BT$.  
The overall numerical coefficient in this linear scaling law was computed
analytically by taking advantage of the near horizon AdS$_3 \times \bR^2$ factor that emerges
in this regime.  On the CFT side, the low temperature physics was seen to be controlled by a
gas of fermions arising from the lowest Landau level of the 4-dimensional gauge theory in 
the presence of a magnetic field.  

\sm

The general case of nonzero $B$ and $\rho$ was studied in \cite{D'Hoker:2009bc},
and the low temperature thermodynamics were found to depend crucially on  
the value of the Chern-Simons coupling~$k$.  This was characterized in terms 
of flows in parameter space as the temperature was lowered.  
These flows head towards three distinct fixed points, depending on the value of~$k$. 
For $k<1$,  the geometry near the horizon can be thought of as  $AdS_2 \times \bR^3$ deformed
by the presence of the magnetic field.  The entropy density was found to be non-vanishing at $T=0$.  
Precisely at $k=1$, the solutions support a near-horizon {\it warped} 
AdS$_3 \times \bR^2$ factor (the warped solutions were studied in the context of topologically massive 
gravity in  \cite{Anninos:2008fx,Compere:2009zj}), and the entropy  density is again  non-vanishing at 
extremality.  For $k>1$ (which includes the 
supersymmetric value $k=2/\sqrt{3}$), the presence of a moderate strength (or larger)  magnetic field 
was seen to lead to a precipitous drop in the entropy density as the temperature was lowered.  
However, our numerics were found to break down in the combined regions of very low temperatures and
small magnetic fields,
presumably as a result of one of our choices of gauge and, as a result,  our analysis 
stopped short of obtaining reliable data for the entropy density at low $T$ over the full
range of magnetic fields.

\subsection{Summary of results}

In the present paper, we shall apply a simple remedy to the gauge choice problem 
which plagued the low temperature numerical work of \cite{D'Hoker:2009bc},
and carry out high-precision numerical analyses down to ultra-low temperatures, 
for a wide range of values of $B^3/\rho^2$. 

\sm

\begin{figure}[h!]
\begin{centering} 
\includegraphics[scale=.6]{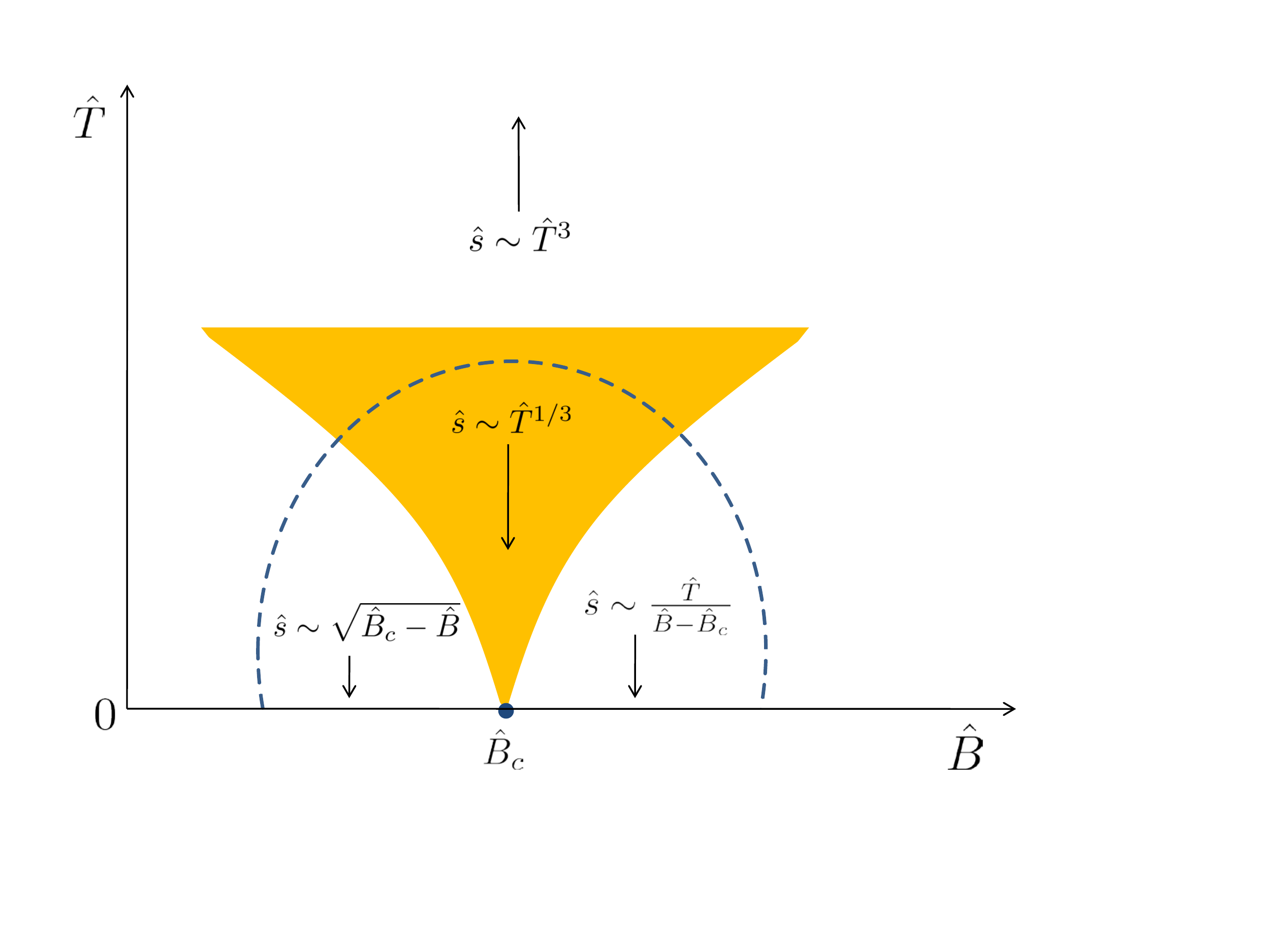}
\caption{Schematic phase diagram illustrating the various  behaviors of the entropy density 
versus temperature and magnetic field. The region inside the dotted line is controlled by the 
quantum critical point at $(\Th=0,\Bh=\Bh_c)$, and  the entropy density  can be 
expressed in terms of a single scaling function $f$ of $(\hat B - \hat B_c)/T^{2/3}$. 
We move around inside this region by changing the temperature $\Th$ and the relevant coupling 
$\Bh-\Bh_c$.   The boundary of the region is defined to be where  irrelevant operators
become important.  The yellow region denotes a regime where temperature is the largest 
energy scale, corresponding to  the argument of the scaling function $f$ being small.  Outside
the yellow region the low temperature behavior of the entropy density, for fixed $\hat B$, 
is either constant or linear in $\hat T$, 
depending on whether the quantum critical point is approached from below or 
from above $\hat B_c$ as $\Th\rightarrow 0$.  }
\label{phase}
\end{centering}
\end{figure}

By virtue of scale invariance, only dimensionless combinations of quantities, 
such as $B^3/\rho^2$, afford any intrinsic physical meaning. Thus, we shall introduce 
normalized, dimensionless, magnetic field $\hat B$,
temperature $\hat T$,  and entropy density $\hat s$  via the following relations, 
\bea 
\label{Tsf}
\Bh \equiv {B\over \rho^{2/3}}
\hskip 0.7in 
 \hat T \equiv { T \over (B^3 + \rho^2)^{1/6}}
\hskip 0.7in 
\hat s \equiv { s \over (B^3 + \rho^2)^{1/2}}
\eea
The results are  displayed schematically  
in Figure \ref{phase} and summarized below.  %
%
\begin{enumerate}
\item Our main result is that a  continuous  quantum phase transition occurs at  
\bea
\label{Bcrit}
\Bh_c  =0.4994240 ~ \pm ~ 0.0000007
\eea 
\item  
For $\Bh>\Bh_c$, the entropy density goes to zero linearly with temperature, $\hat s \sim \hat T$, 
reflecting the same Fermi liquid physics that was seen in the large $\Bh$ limit.
Here, however, the linear behavior is occurring at finite charge density where the result does not 
follow from conformal invariance.   Note that the specific heat $C$, at constant $\Bh$,
has the same behavior as the entropy density, since $C = \hat T\partial \hat s /\partial \hat T \sim \hat T$.  
\item 
On approaching $\Bh_c$ from the large $\hat B$ side, the coefficient of the linear
term in the $\hat s$ versus $\hat T$ relation is found to diverge according to, 
\bea
{\hat s \over \hat T}  \sim {1 \over (\Bh - \Bh_c)^\sigma} \hskip 1in \sigma = 1.003 ~ \pm ~ 0.005
\eea
This signals a breakdown of Fermi liquid behavior. 
\item 
Approaching the quantum critical point along the temperature axis at fixed $\Bh=\hat B_c$, 
the entropy   density exhibits a new power law scaling,
\bea
\label{alpha}
\hat s \sim \hat T^\alpha \hskip 1in \alpha = 0.335 ~ \pm ~ 0.005
\eea
\item 
For $\Bh<\Bh_c$ the zero temperature entropy  density is found to be nonzero.  Near 
$\Bh_c$  it obeys the scaling,
\bea
\hat s \sim (\Bh_c -\Bh_c)^\tau \hskip 1in \tau = 0.500 ~\pm ~ 0.001
\eea
\item 
The entropy  density in the vicinity of  the fixed point can be expressed in terms
of a scaling function $f$ as,
\bea
\hat s = \hat T^{1/3} f\left( {\Bh-\Bh_c \over \hat T^{2/3} }\right)
\eea
This is to be contrasted with regions far from the critical point, where the entropy  
density  is a nontrivial
function of two dimensionless combinations of $\hat \rho$, $\hat B$ and $\hat T$.

\end{enumerate}

From our results, we can infer that the quantum critical theory lives in $1+1$ spacetime dimensions
(that is, there are no long range correlations in the remaining 2  spatial directions along the boundary), 
has dynamical critical exponent $z=3$ (since the value found numerically for $\alpha$ is consistent 
with $\alpha =1/3$), and has a relevant operator with scaling dimension~$2$.   The relevant operator 
corresponds to a change of $\Bh$ away from $\Bh_c$.  

\subsection{Comparison with known quantum critical systems}

It is illuminating to place these results within the context of known quantum critical systems.
Although thermodynamic quantities such as the specific heat and magnetization 
behave in a non-analytic fashion across the phase transition,   
we note that there is no change of symmetry associated with the transition.
A finite temperature example of such behavior is the liquid-gas transition in water.  
A zero temperature version
is a metamagnetic quantum critical point \cite{Millis02,RPMMG}, whose behavior closely parallels our system  (a finite temperature version of a  holographic metamagnetic phase transition in the D4-D8 system was  studied in \cite{Lippert}).  
Metamagnetism refers to a sharp change in the  magnetization of a material as an 
external  magnetic field is tuned through some {\sl nonzero} value.  
At finite temperature, metamagnetic transitions typically consist
of first order lines  terminating at a second order critical point, as in the liquid-vapor case.  If
the critical point can be brought to zero temperature by adjusting some control parameter, one
obtains a metamagnetic quantum critical point.  

\sm

The standard approach to such a quantum critical point is based on the Hertz/Millis 
theory \cite{hertz,millis93,LRMW}, for which the effective action in momentum space is, 
\bea
S = \int\! d\omega d^dk \left( {|\omega| \over |k| } + k^2 + (\Bh-\Bh_c) \right) |\phi(\omega,k)|^2 + \cdots
\label{HM}
\eea
The real  bosonic field $\phi$ represents the local  magnetization, and the above action can be obtained
by integrating out fermions at 1-loop.   Under scale transformations acting as $k\rightarrow \lambda k$,
we see that $\omega$ should be assigned  scale dimension 3, and hence $z=3$.   Similarly,
$\Bh-\Bh_c$ has scale dimensions $2$.     These assignments match what we found for our system, which
furthermore corresponds to $d=1$, since the Landau level quantization only allows low energy
modes to propagate parallel to $B$.  

\sm

The action  (\ref{HM}) is only meant to be applied for $\Bh>\Bh_c$.  
Indeed, the behavior of our system in the region $\Bh<\Bh_c$, with its nonzero ground 
state entropy density, cannot be described by this action alone. 

\sm 

Metamagnetic quantum criticality in the compound Sr$_3$Ru$_2$O$_7$ has been the subject
of extensive experimental investigation in the past few years (see \cite{RPMMG} and references therein), 
and we will comment on this connection in Section 3.8. 

\sm

A large  number of other AdS/CFT examples undergoing phase transitions have been studied in the
literature, both continuous and discontinuous, and at finite and zero temperature; for example 
 \cite{Witten:1998zw,Chamblin:1999tk,Parnachev:2006dn,Mateos:2006nu,Davis:2007ka,Hartnoll:2008vx,
Davis:2008nv,Evans:2010iy,Jensen:2010vd}.   However, our
setup seems particularly attractive and is nicely  related to real experimental systems.   In particular, 
unlike other examples of quantum phase transitions we do not have to add any extra ingredients
in the way of scalar fields or probe branes.  We employ only a metric and an Abelian gauge field,
with an Einstein-Maxwell-Chern-Simons action that is known to describe {\sl all}  supersymmetric
AdS$_5$ theories related by compactification of Type IIB or M-theory 
\cite{Buchel:2006gb,Gauntlett:2006ai,Gauntlett:2007ma}.  
Thus our framework is both simple and universal.

\sm

The remainder of this paper is organized as follows. In Section 2, we spell out the set-up
of the holographic calculations, including the specification of initial data at the horizon, 
asymptotic data at the $AdS_5$ boundary, and the construction of regular gauge choices.
In Section 3, we present the results of high-precision numerical solutions to the 
reduced Einstein-Maxwell-Chern-Simons equations, identify the quantum critical point, and
critical behavior in its vicinity. We also compare our results with those from the effective
Hertz/Millis theory, and comment on the connection with metamagnetic quantum criticality
observed in real materials like Sr$_3$Ru$_2$O$_7$. A discussion of the results and open 
avenues for future research is given in Section 4.

\section{Holographic calculations}
\setcounter{equation}{0}

In this section we spell out some of the technical details of our  computations.  
Results of the numerical calculations will be presented in  section 3, and the impatient 
reader may wish to jump there. 

\sm

The starting point for our holographic calculations is 5-dimensional Einstein-Maxwell
theory with a Chern-Simons term.  Throughout this paper, the Chern-Simons coefficient
$k$ will be considered fixed at its supersymmetric value $k=2/\sqrt{3}$.  A detailed discussion of the action, including 
boundary terms needed for holographic renormalization, and the construction of 
the boundary current and stress tensor may be found in \cite{D'Hoker:2009bc}.
Here, we shall limit our discussion to the field equations and the asymptotic behavior
of the fields near the horizon, and at the asymptotic $AdS_5$ boundary.
The Einstein-Maxwell field equations are given by $dF=0$ and,
\bea
0 & = & d *F + k F \wedge F
\no \\
R_{MN} & = & 4 g_{MN} + { 1 \over 3} F^{PQ} F_{PQ} ~ g_{MN} - 2 F_{MP} F_N {}^P
\eea
Uniformity and constancy in time of the magnetic field $B$ and the charge density $\rho$ 
allow us to restrict to a space-time translation invariant  Ansatz, given by,
\bea
\label{ansatz}
F & = & E dr \wedge dt + B dx_1 \wedge dx_2 + P dx_3 \wedge dr
\no \\
ds^2 & = & U^{-1} dr^2 - U dt^2 + e^{2 V} \left ( dx_1^2 + dx_2^2 \right ) 
+ e^{2W} \left ( dx_3 + C dt \right )^2
\eea
The functions $E,P,U,V,W,C$ depend only on the radial coordinate $r$,
while the magnetic field  $B$ is constant by the Bianchi identity. A gauge 
choice has been made here for the coordinate $r$ in order to put the 
Ansatz in {\sl canonical form} with matching coefficients of its first two terms in $ds^2$.
The reduced field  equations were given in \cite{D'Hoker:2009bc}. 

\subsection{Data at the horizon}

The reduced field equations are to be solved subject to regularity conditions
at the horizon and at the asymptotic $AdS_5$ boundary $r=\infty$. For the 
purpose of numerical analysis, it will be convenient to parametrize solutions 
in terms of data at the horizon which satisfy the regularity conditions at 
the horizon. Regularity of the full solution, including at the asymptotic $AdS_5$ 
boundary, must then be verified numerically for each set of data. 

\sm

We begin by spelling out the data at the horizon. (This discussion will parallel the
one presented in \cite{D'Hoker:2009bc}, but there will be important differences 
motivated by the need to remedy the gauge choice problems alluded to in the Introduction.) 
The (outer) horizon at $r=r_+$, and the Hawking temperature $T$ are defined by,
\bea
U(r_+)=0 \hskip 1in 4 \pi T = U'(r_+)
\eea
Our numerical analysis will always be carried out at $T\not=0$, even though 
$T$ may become very small; thus, we are free to rescale $t$, and set $U'(r_+)=1$.
By rescaling also $x_1, x_2, x_3$, we may set $V(r_+)=W(r_+)=0$. Invariance of the 
Ansatz under {\sl $\alpha$-symmetry}, (under which $x_3 \to x_3 - \a t$, 
$C\to C + \a$, and $E \to E - \a P$), allows us to set $C(r_+)=0$. With these 
choices, the fields at the horizon take the form,
\bea
F_H & = & q \, dr \wedge dt + b \, dx_1 \wedge dx_2 + p \, dx_3 \wedge dr
\no \\
ds_H^2 & = & dx_1^2 + dx_2^2 + dx_3^2
\eea
The reduced field equations of \cite{D'Hoker:2009bc} for $E,P,U,V,W,C$, 
combined with the requirement of regularity  at the horizon, 
dictate certain relations amongst the remaining data  at the horizon, 
namely $q,b,p, V'(r_+)$, $W'(r_+)$, and $C'(r_+)$. 
They are given by (for $U'(r_+)=1$),
\bea
\label{hordata}
0 & = & p - q\left ( C'(r_+) - 2 k b \right )  
\no \\
3V'(r_+) & = & 12 - 2 q^2 - 4 b^2
\no \\
6 W'(r_+) & = & 24 - 4 (q^2 - b^2) - 3 C'(r_+)^2
\eea
Given $(b,q)$, the data $p$ and $ C'(r_+)$ are related to one another by the 
first equation of (\ref{hordata}). Keeping either $p$ or $C'(r_+)$ as an independent 
free parameter, the remaining initial data $V'(r_+)$ and $W'(r_+)$ are uniquely 
determined by the  last two equations of (\ref{hordata}).

\subsection{Gauge fixing and regularity}

The parametrization  of the horizon data presented above is more general 
than the one given in \cite{D'Hoker:2009bc}, since here the parameter $p$ is kept
unspecified, while $p$ was set to 0 in \cite{D'Hoker:2009bc}. The argument invoked
to set $p=0$ was the covariance of the Ansatz of (\ref{ansatz})
under boosts in the $x_3$-direction. Under a boost by velocity $\beta$, the space-time
coordinates transform as usual, 
$t \to \tilde t = \gamma (t - \beta x_3)$ and $x_3 \to \tilde x_3 = \gamma  (x_3 - \beta t)$
with $\gamma  ^2 (1 - \beta ^2)=1$,  and where $\beta$ cannot exceed the speed of 
light,  $|\beta| < 1$. The transformation under boosts of the functions 
$(E,P,U,V,W,C)\to (\tilde E, \tilde P, \tilde U, \tilde V, \tilde W, \tilde C)$, however, 
must be accompanied by a transformation
of the holographic coordinate $r \to \tilde r$, which is required to restore the boosted 
Ansatz back to the {\sl canonical form} of (\ref{ansatz}). As a result, the Maxwell
fields transform as follows,
\bea
\tilde E (\tilde r) d\tilde r & = & \gamma  \Big ( E(r) - \beta P(r) \Big ) dr
\no \\
\tilde P (\tilde r) d\tilde r & = & \gamma  \Big ( P(r) - \beta E(r) \Big ) dr
\eea
where $\tilde U(\tilde r)^{-1} d \tilde r = U(r)^{-1} dr$. The ratio 
$p/q$ transforms as $\tilde p/ \tilde q = (p/q - \beta )/(1 - \beta p/q)$.
Clearly, provided $q^2>p^2$, a boost by $\beta = p/q$ may be used to set 
$\tilde p$ to zero.  The problem, however,  is that the 
coordinate transformation  $r \to \tilde r$ required to accompany this boost may be 
singular on some of the functions $E,P,U,V,W,C$. For example, the transformation
law under a boost by velocity $\beta$ of the function $C$ is given by,
\bea
\label{cc}
C(r) \to \tilde C(\tilde r) = { (C(r)+\beta) (1+ \beta C(r)) e^{2 W(r)} - \beta U(r)
\over (1 + \beta C(r))^2 e^{2 W(r)} - \beta ^2 U(r)}
\eea
We have verified numerically that, in the region of low $T$ where our numerics 
were found to break down in \cite{D'Hoker:2009bc},  the denominator in (\ref{cc}) 
indeed crosses zero as $r$ is increased away from the horizon at $r_+$,
thereby rendering this boost transformation singular.

\sm

The remedy to this problem is simple: the parameter $p$ should be left unspecified, 
thereby eliminating the need to perform boost transformations on the fields.
It will be convenient to parametrize solutions by the values  $b,q,C'(r_+)$,
so that the remaining initial data at the horizon, $p, V'(r_+)$, and $W'(r_+)$ are 
uniquely determined by (\ref{hordata}). Not every assignment of $b,q,C'(r_+)$
will produce a regular solution. Also, two regular solutions may be related
to one another by a {\sl regular} boost, and thus be physically equivalent.
The parameter space of all regular solutions may be described as follows.
To every pair $(b,q)$ we assign the maximal interval $\Gamma (b,q)$ on the 
real line such that for 
every value of $C'(r_+) \in \Gamma (b,q)$, the solution specified by the parameters 
$b,q,C'(r_+)$ is regular. The end points of the interval $\Gamma (b,q)$ 
correspond to the limits where the velocity of the solution tends to the speed of light.
When computing boost invariant 
physical quantities, such as the magnetic field $B$, and the rest frame temperature $T$,
charge density $\rho$,  and entropy density $s$, the datum $C'(r_+)$ may be chosen to be any 
value in the interval  $\Gamma (b,q)$. 

\subsection{Data at  the asymptotic $AdS_5$ boundary $r \to \infty$}

For regular solutions the functions $U,V,W,E,P,C$ have the 
following asymptotic behavior as $r \to \infty$, (keeping only leading contributions), 
\bea
\label{asymdata}
U \sim r^2 \hskip 0.85in & \qquad e^{2 V} \sim v r^2  & \qquad e^{2W} \sim w r^2 
\no \\
E \sim  (e_3 - c_0 p_3)r^{-3} & \qquad P \sim p_3 r^{-3} & \qquad 
C \sim c_0
\eea 
In these coordinates, the conformal boundary metric is 
$-dt^2 + v (dx_1^2+dx_2^2) + w (dx_3+c_0 dt)^2$,
and the solution's velocity is $\sqrt{w} ~ c_0$. Rescaling $x_1,x_2$ by $\sqrt{v}$, and
$x_3$ by $\sqrt{w}$, while combining an $\alpha$-transformation with a boost by 
a velocity $\beta = - \sqrt{w} c_0$, 
restores the coordinates to the standard Minkowski metric, and yields the following 
expressions for the physical magnetic field $B$, temperature $T$, 
charge density $\rho$, and entropy density $s$ 
{\sl in the rest frame of the solution},\footnote{In the CFT, the rest frame 
corresponds to a statistical
ensemble weighted by the Boltzmann factor $e^{-(H-\mu Q)/T}$.   
Boosting produces additional chemical potentials multiplying momenta 
and currents.}  (see \cite{D'Hoker:2009bc} for derivations),
\bea
B={ b \over v} \hskip 0.6in T = { \gamma _c \over 4 \pi} \hskip 0.6in 
\rho = \gamma _c ( e_3 - c_0 p_3) \hskip 0.6in 
s = { 1 \over 4 v \gamma _c \sqrt{w}}
\label{phys}
\eea
Here, $\gamma _c^2 = (1-wc_0^2)^{-1}$, and the normalized entropy density $s$ is defined as 
$s = G_5 S/{\rm Vol}$, where $S$ is the total entropy in volume Vol, and 
$G_5$ is the 5-dimensional Newton constant.

\sm

By virtue   of scale invariance, only dimensionless combinations of quantities, such as 
$B^3/\rho^2$,  afford any intrinsic physical meaning. Thus, the results on phase transitions 
and associated critical exponents at critical points,  aimed for in this paper, will all be 
derived by evaluating the dependence between the dimensionless physical quantities $\hat s$
and $\hat T$, for various (fixed) values of $\Bh$, which were all defined in (\ref{Tsf}).

\subsection{Numerical fine points}

The initial data for any regular solution is a pair $(b,q)$ and a value 
$C'(r_+)\in \Gamma (b,q)$, where the interval $\Gamma (b,q)$ was defined so
that the data $b,q,C'(r_+)$ produce a regular solution. For certain values of 
$(b,q)$, the interval $\Gamma (b,q)$ may be empty. For $k =2/\sqrt{3}$, numerical 
analysis yields the following bound, 
\bea
\Gamma (b,q) = \emptyset \hskip 0.8in {\rm when} \quad q^2 + 2 b^2 >6
\eea
The precise form of the critical curve $\cC$ in the $(b,q)$-plane inside of which 
$\Gamma (b,q)\not= \emptyset$ is not known analytically, but may be obtained 
numerically, as was done in \cite{D'Hoker:2009bc}.  

\sm

Away from the low temperature regime the interval $\Gamma(b,q)$ is sufficiently large that it
is possible, at least for  $k=2/\sqrt{3}$, to make a single uniform choice for $C'(r_+)$ and still
cover most of the parameter space.  A convenient choice is  $C'(r_+) = 2 b$.  
However, as the temperature is lowered $\Gamma(b,q)$ shrinks, and one is forced to tune
$C'(r_+)$ to greater precision.  This effect becomes especially pronounced in the region
of low magnetic field where a ground state entropy develops.  However, by specifying
more general tunings for $C'(r_+)$ as we vary $(b,q)$ we are able to fully  cover the low 
temperature region.

\sm

The approach to ultra-low temperatures, which is needed in various parts of our 
numerical work, requires a high degree of fine-tuning of the horizon initial data 
$(b,q)$ and the gauge choice $C'(r_+)$. It also requires evaluating the asymptotic
data, such as $v,w,c_0,e_3$ and $p_3$ at large values of $r$, which we typically 
have taken to range from $10^{15} $ to $10^{20}$. With such high degrees of fine-tuning,
and extended ranges of integration, the issue of numerical accuracy and numerical 
stability of the calculations becomes of utmost importance. Our calculations were 
performed with 15 to 20 digits accuracy, and the ODEs were solved with absolute 
and relative error tolerances ranging from $10^{-10} $ to $10^{-13}$ for the lowest
temperatures. Stability of the results was checked versus changing the asymptotic
value of $r$, the absolute and relative error tolerances, and the number of digits.

\section{Numerical results}
\setcounter{equation}{0}

In this section, we present our numerical results, organized 
as a function of the magnitude of the magnetic field $\hat B$, starting at large $\hat B$.

\subsection{Large $\Bh$ regime} 

In \cite{D'Hoker:2009bc}  the case $\Bh \approx .53$ (quoted there as  $\Bh^3 \approx .15$) 
was studied at the supersymmetric value 
$k=2/\sqrt{3}$.  As displayed in Figure 3 of \cite{D'Hoker:2009bc} the low temperature 
entropy density was seen to drop well below its $\Bh=0$ value, and appeared 
to be heading towards zero.
However, the numerics broke down at temperature $\Th \approx .02$, and so did not allow
for an exploration of ultra-low temperatures.   
By adjusting $C'(r_+)$ as we lower the temperature, we can do much better, as shown in 
Figure \ref{B53} below.   The entropy density clearly vanishes at zero temperature.  
A linear behavior, $\sh \sim \Th$ is
evident in the approach to zero temperature, as shown in the right panel  of Figure \ref{B53}.  
\begin{figure}[h!]
\begin{centering}
\includegraphics[scale=.7]{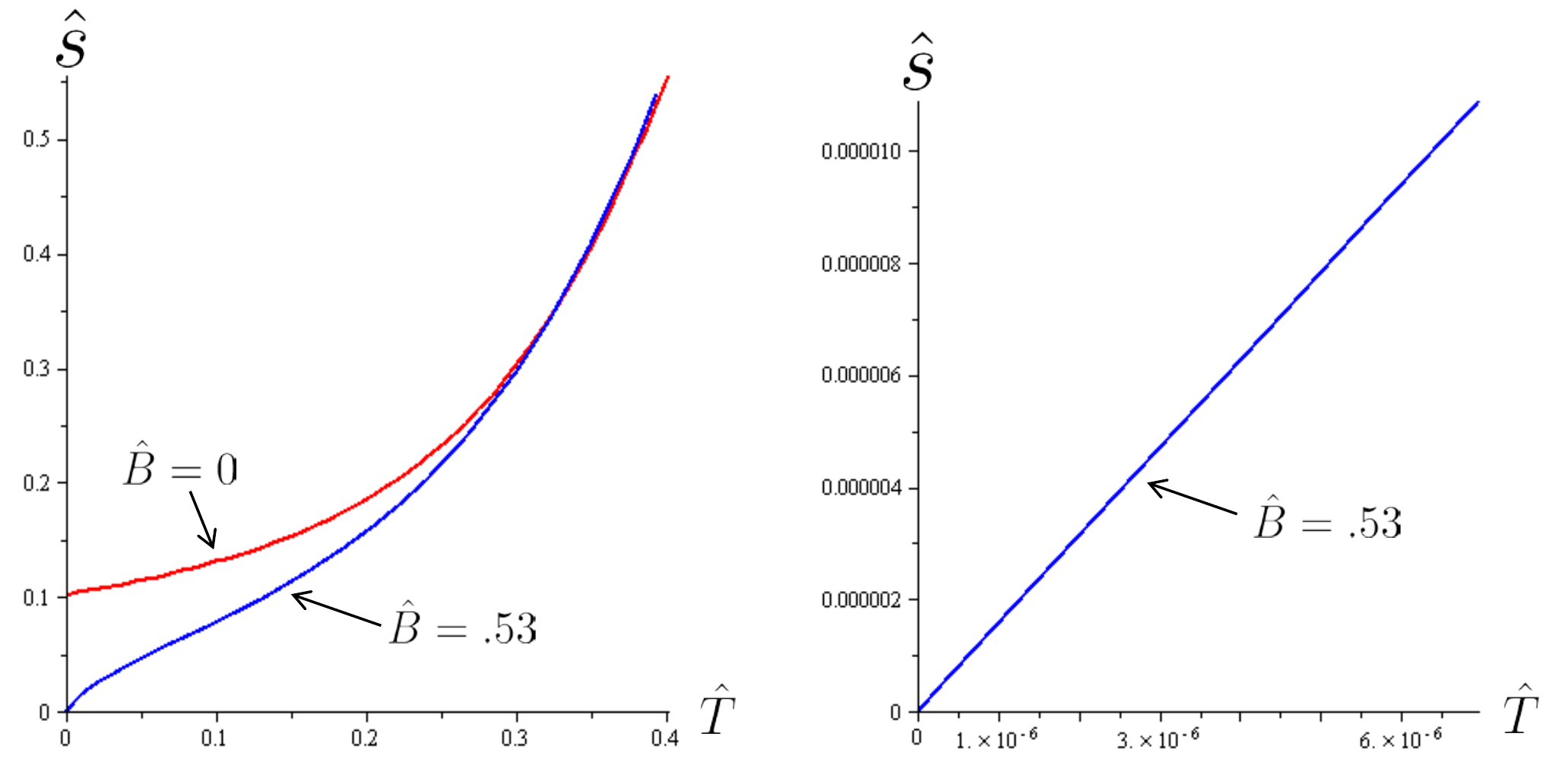}
\caption{Plot of entropy versus temperature.  On the left we compare $\Bh=.53$ to $\Bh=0$;
this plot is an improved version of Fig. 3 in \cite{D'Hoker:2009bc}.  On the right we exhibit
the linear $\sh \sim \Th$ low temperature behavior.  }
\label{B53}
\end{centering} 
\end{figure}
By repeating this analysis for other {\sl sufficiently large}  values of $\Bh$, we find similar behavior:
the entropy density  vanishes at zero temperature, and does so linearly at very low temperatures.  
This linear
behavior is characteristic of Fermi liquids. The textbook intuitive explanation for the linear dependence 
is that at low temperature the smoothed out step function form of the  Fermi-Dirac distribution implies
that only electrons within  energy $k_B T$ of the Fermi energy contribute.  We indeed expect our system
to be described by a theory of light fermions in this large $\Bh$, low $\Th$ regime, since the magnetic 
field  raises up all energies  except for the lowest fermion Landau level.    Our numerical results are
a pleasing confirmation of this intuition. 

\sm 

In the $\Bh\rightarrow \infty$ limit it was found in \cite{D'Hoker:2009mm} that a near horizon
AdS$_3 \times \bf R^2$ factor emerged at low energies, and the resulting $1+1$ dimensional
conformal invariance could be used to explain the $\sh \sim \Th$ behavior.   This is no longer
true away from this limit, as the charge density $\rho$ introduces an additional scale into the problem, and
from the numerics we can see that the near horizon geometry becomes deformed away from
 AdS$_3 \times \bf R^2$. 

\sm

Charged black hole  solutions involving scalar fields whose entropy densities similarly go to zero at extremality have
been studied in \cite{Horowitz:2009ij,Goldstein:2009cv,Cadoni:2009xm,Gauntlett:2009bh}.  

\subsection{Approaching $\Bh_c$}

\begin{figure}[h!]
\begin{centering}
\includegraphics[scale=.52]{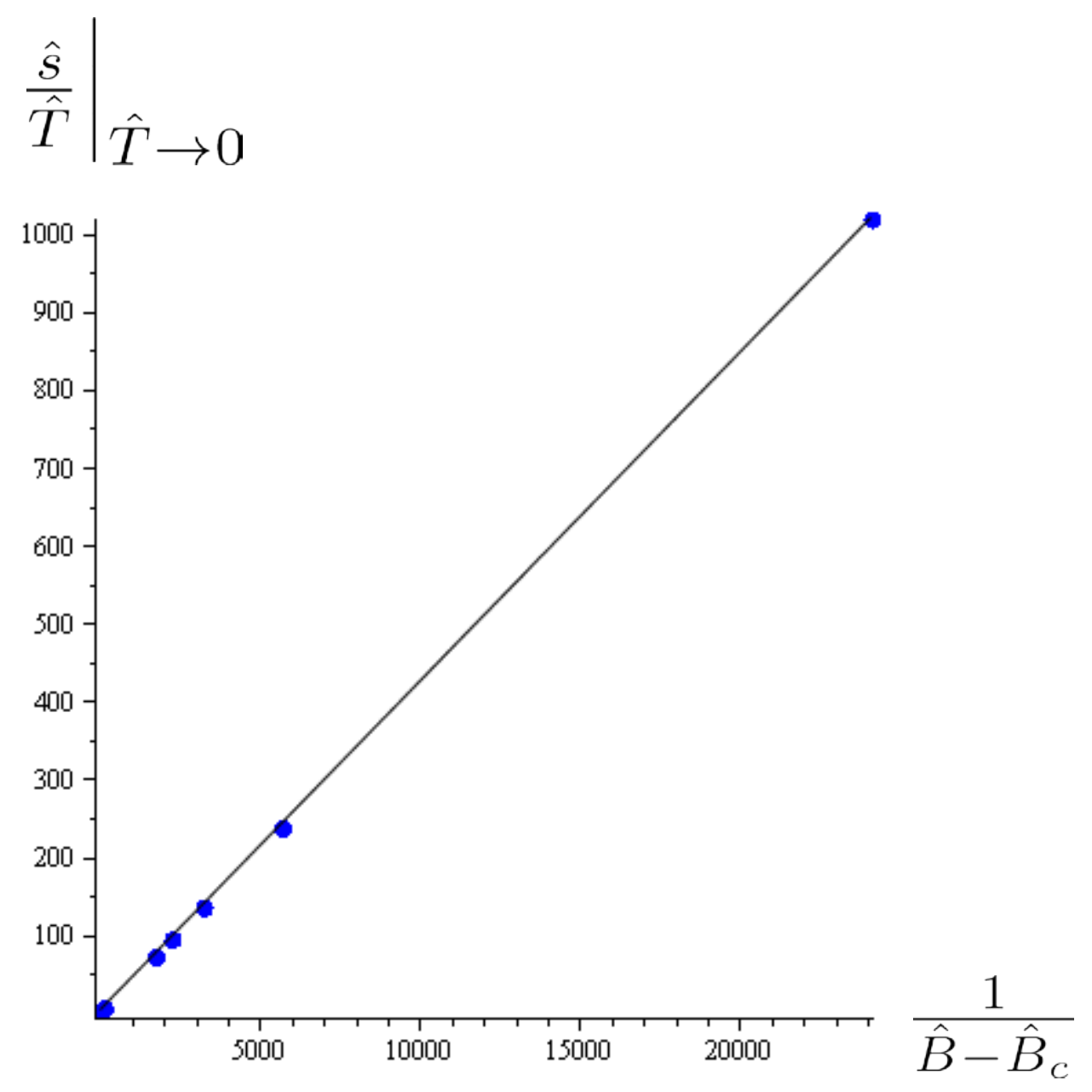}
\caption{Plot showing the divergence of  $\sh/\Th$ near $\hat B_c$ at
low temperatures.   The straight line through the data points is included to guide the eye.  }
\label{sdiv}
\end{centering}
\end{figure}

Experimentally, a breakdown of Fermi liquid behavior upon tuning an external magnetic field 
can often be seen in a divergence of the specific heat coefficient, defined as,
\bea
\gamma = {C\over \hat T} 
\eea
Since $C=\hat T {\partial \hat s/\partial \hat T}$, we can equivalently write 
$\gamma = \hat s/\hat T$ at low
temperatures.    In our system, $\gamma $ stabilizes at a constant value for large 
$\Bh$, but is seen to diverge at a critical value $\Bh=\Bh_c$. Numerically, the critical
value $\hat B_c$ is found to be bounded as follows,
\bea
\label{Bcubed}
0.124568 < \hat B_c ^3 < 0.124569
\eea
which results in the the value quoted already in (\ref{Bcrit}), namely 
$\hat B_c = 0.4994240 \, \pm \, 0.0000007$. To characterize the divergence we 
display a  plot of 
$\sh / \Th $ versus $1/(\Bh-\Bh_c)$, as $\Th \to 0$ in Figure \ref{sdiv}.  
The straight line shows that  the low temperature entropy behaves as 
\bea
{\sh \over \hat T} \sim  {1 \over \Bh-\Bh_c}
\eea
as $\Th\rightarrow 0$ for fixed $\Bh$ near (but larger than) $\Bh_c$. 
Upon approaching $\Bh_c$ we find that the linear regime with $\hat s \sim \Th$ 
is confined to an ever smaller low temperature window. 

\subsection{Scaling at the critical magnetic field $\hat B = \hat B_c$}

\begin{figure}[h!]
\begin{centering}
\includegraphics[scale=.5]{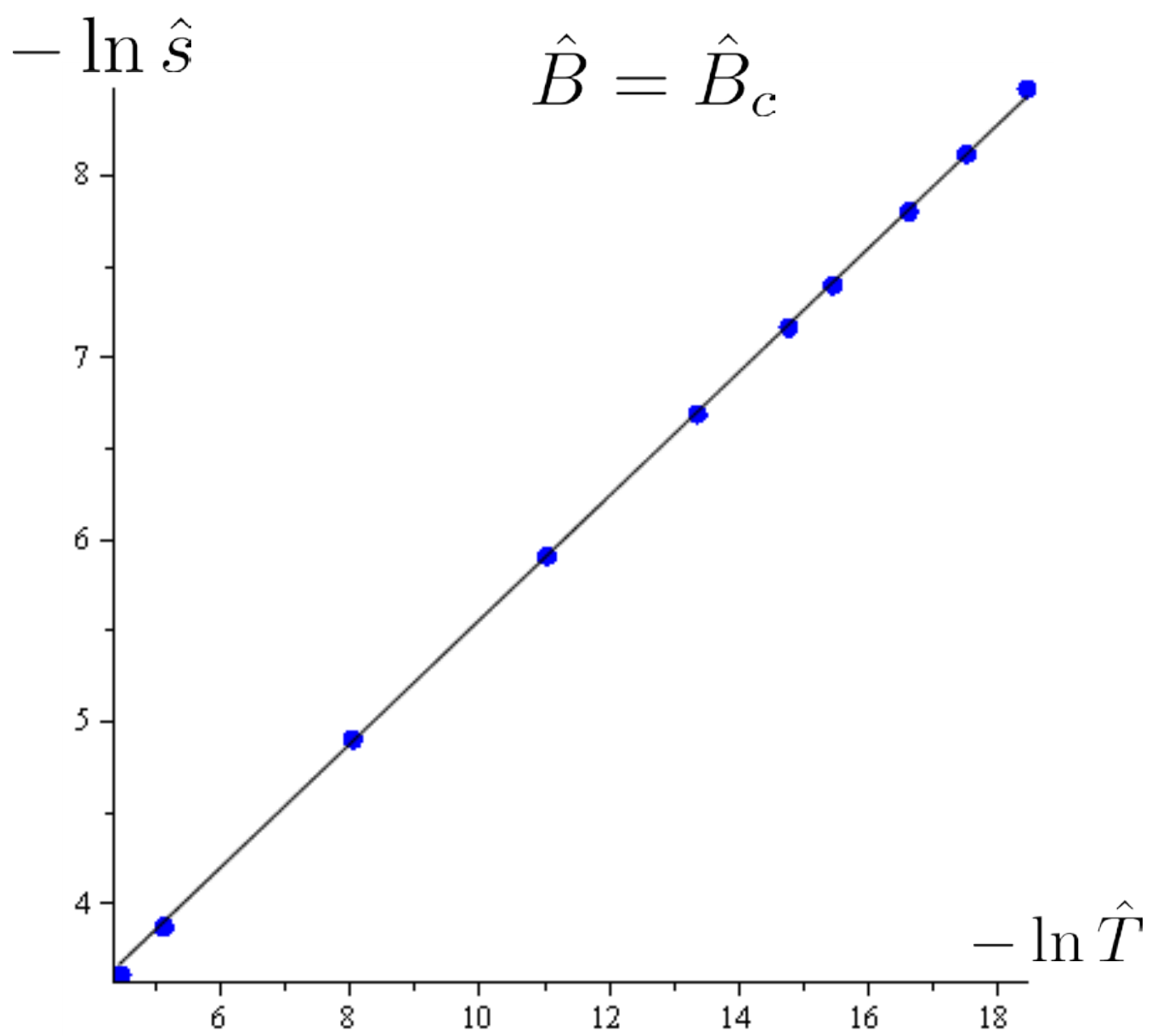}
\caption{Plot showing $\sh \sim \Th^{1/3}$ scaling behavior at $\Bh=\Bh_c$.
The straight line through the data points is included to guide the eye, 
and has slope $1/3$, consistent with  (\ref{alpha}).}
\label{T13}
\end{centering}
\end{figure}

Next we set $\Bh =\Bh_c$ and again study the low temperature behavior of the entropy density.   
We find a new scaling law, $\sh \sim \Th^{1/3}$, which numerically extends over at least four orders 
of magnitude in temperature, $10^{-8} < T < 6 \times 10^{-3}$,  as shown in Figures \ref{T13}.   
This nontrivial power law manifestly  represents  non-Fermi liquid behavior, analogous to what 
is seen in real materials at a quantum critical point.  

\sm

Towards ultra-low temperatures, the numerical behavior of $\hat s$ will ultimately 
turn over to be linear in $\hat T$ (for $\hat B > \hat B_c$) or to be a non-zero constant
for $\hat B < \hat B_c$ (see the next subsection). This deviation from $\hat s \sim \hat T^{1/3}$
scaling  is caused by the fact that the value of $\hat B_c$ is known only numerically from (\ref{Bcrit}),
and we are never able to sit at {\sl precisely} the value of $\Bh$.

\subsection{The quantum critical region and crossover}

The $\hat s \sim \hat T ^{1/3}$ scaling law extends to the vicinity of the critical magnetic 
field $\hat B_c$, but the range over which this scaling law holds shrinks as $|\hat B - \hat B_c|$
is increased. The corresponding behavior of $\hat s$ is illustrated in Figure 5.

\sm

First, on the right panel of Figure 5, $- \ln(\hat s)$ is plotted versus $-\ln (\hat T)$
for $\hat B ^3 = 0.124569$. The $\hat T^{1/3}$ scaling law exhibited in the preceding 
subsection, is clearly recovered here. At sufficiently high temperature, the $\Th^{1/3}$ 
behavior eventually crosses over to the $\Th^3$ dependence controlled by the UV theory.  
This is clearly shown on the right panel of Figure 5, where the cross-over region may be 
identified with the temperature interval $0.02 < T_{\rm cross-over} < 0.5$.

\sm

Second, on the left panel of Figure 5, the flows of $-\ln(\hat s)$ as a function of $-\ln(\hat T)$
at various fixed values of $\hat B$ are shown. Curves a, b, c, d, e, and f clearly exhibit
the turnover from $\hat s \sim \hat T ^{1/3}$ scaling behavior to linear $\hat s \sim \hat T$
behavior at ultra-low temperatures. Given this cross-over behavior, we know that the 
corresponding values of $\hat B$ must all be {\sl above} the critical magnetic field $\hat B_c$,
with value closest to critical corresponding to curve f with $\hat B^3 = 0.124569$. 
Curve g behaves completely differently at ultra-low temperatures, and $\hat s$ is seen
to tend towards a non-zero constant. Given this behavior, we know that the corresponding
value of $\hat B$ must be {\sl below} critical. Combining both results gives
$0.124568< \hat B_c^3 < 0.124569$, as announced in (\ref{Bcubed}), and (\ref{Bcrit}).

\sm

Finally, as $\hat B - \hat B_c$ is increased, the region over which the $\hat s \sim \hat T ^{1/3}$
scaling law holds becomes smaller and smaller, being overtaken by the $\hat s \sim \hat T^3$
law at high $\hat T$ joining the $\hat s \sim \hat T$ law at low $\hat T$, and ultimately 
will have disappeared altogether by the time $\hat B^3=0.2$ is reached (this result is not shown in the 
figures).

\begin{figure}[h!]
\begin{centering}
\includegraphics[scale=.65]{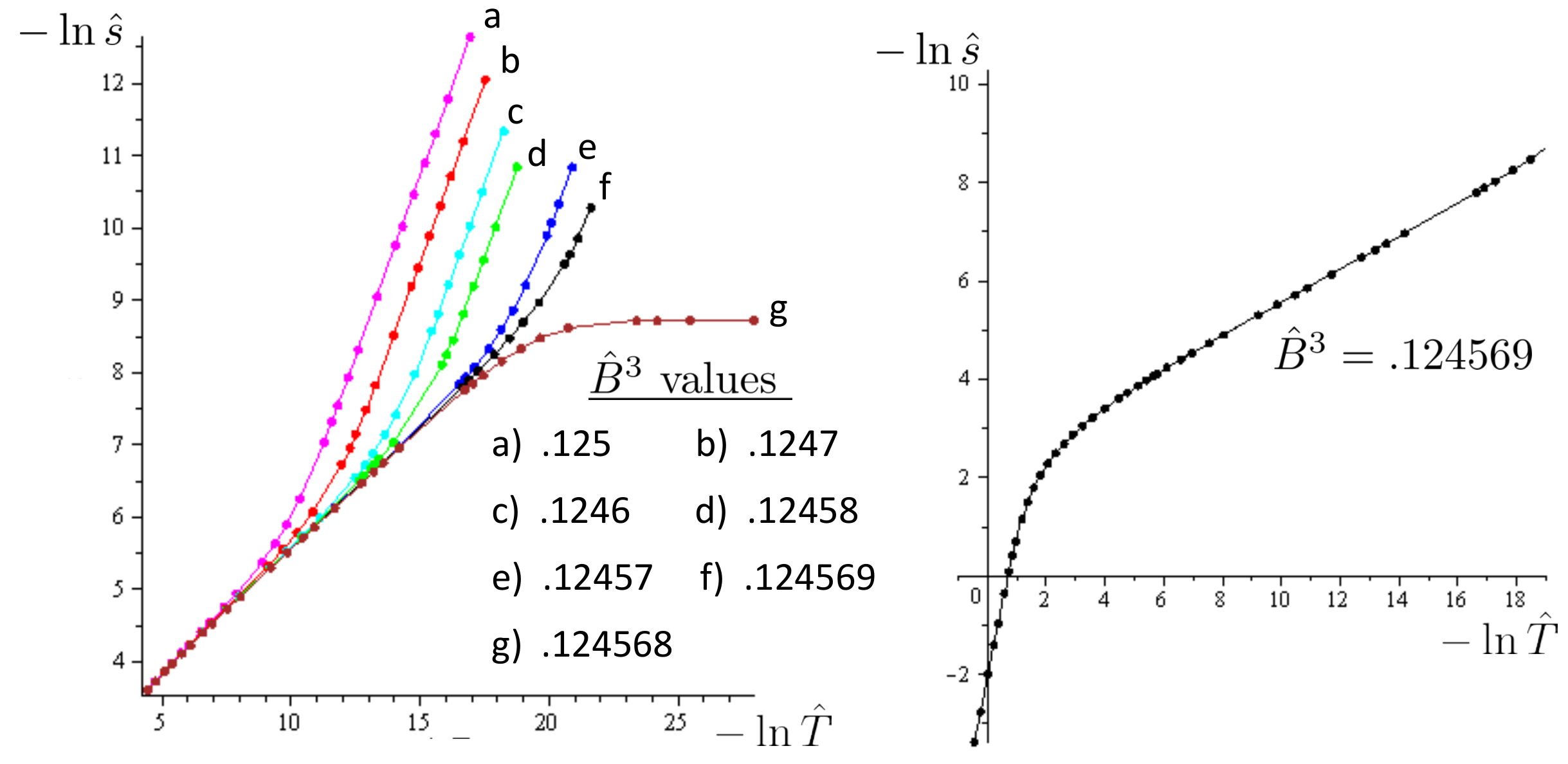}
\caption{The left plot shows the crossover of $\hat s$ for low $\hat T$.
At moderately low temperatures $\hat s$ scales as $\Th^{1/3}$ 
(lower left corner of the plot), while at ultra-low temperatures $\hat s$ scales as $\hat T$
for $\hat B > \hat B_c$ (curves a, b, c, d, e, f), and tends to a non-zero constant for 
$\hat B < \hat B_c$ (curve g). The right plot shows the crossover for $\hat s$ 
from the moderately low temperature $\hat T^{1/3}$ scaling to the high temperature 
$\hat T^3$ behavior. The dots represent numerical data points, while the solid
interpolating lines are included to guide the eye.}
\label{cross}
\end{centering}
\end{figure}

\subsection{Low $\Bh$ region} 

\begin{figure}[h!]
\begin{centering} 
\includegraphics[scale=.5]{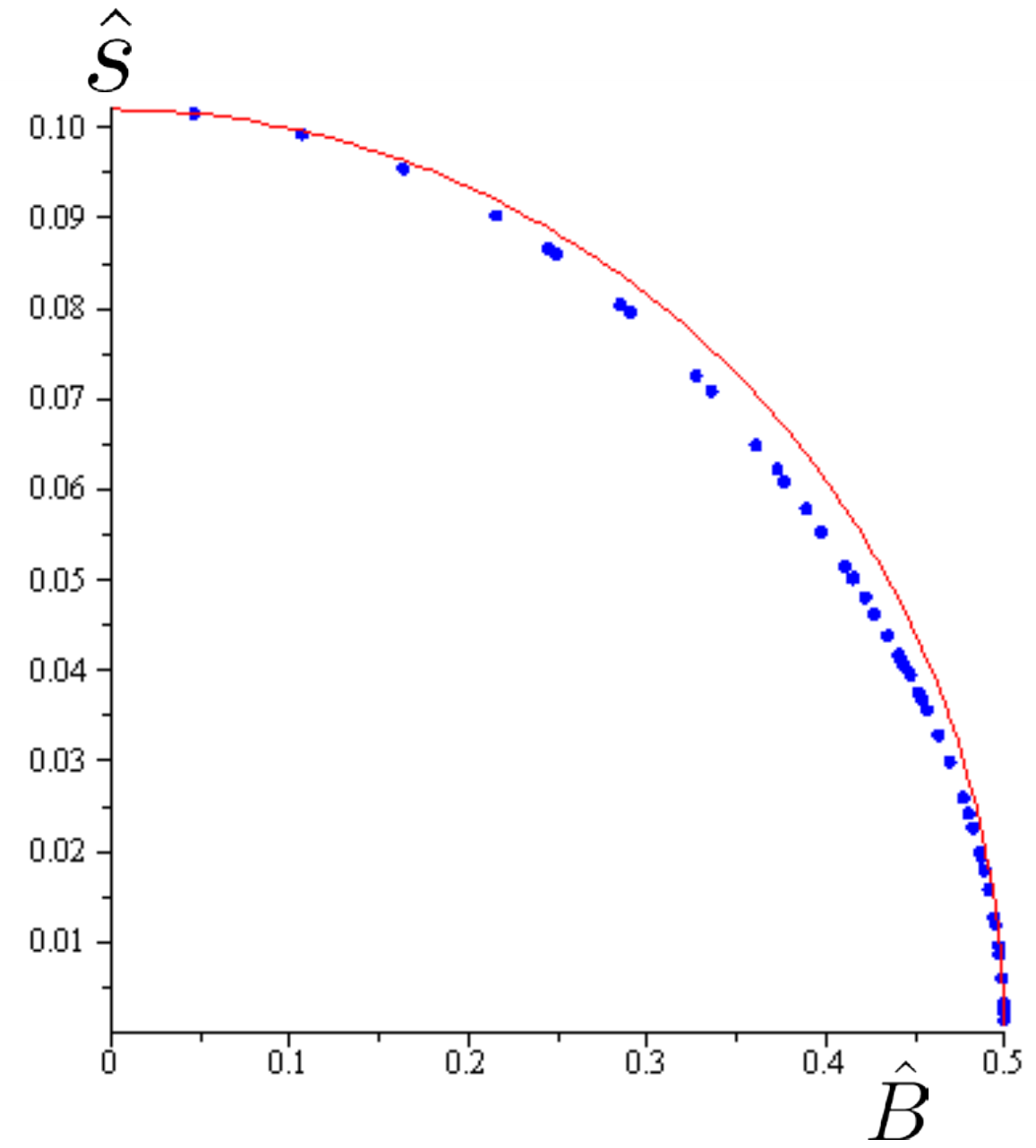}
\caption{Extremal entropy density: blue points represent  results for the entropy density at very 
low temperatures in the region $\Bh<\Bh_c$.  The red curve is a plot of the function 
${1\over 4\sqrt{6}}\sqrt{1-\left({\Bh\over \Bh_c}\right)^2}$ }
\label{extremalS}
\end{centering}
\end{figure}

Decreasing the magnetic field  further, namely $\Bh < \Bh_c$, we find a completely different low
temperature behavior: the entropy now stabilizes to a nonzero value at $\Th=0$.  We plot this
limiting value in Figure \ref{extremalS}.    In \cite{D'Hoker:2009bc} we speculated that an infinitesimally
small magnetic field might be sufficient to remove the ground state entropy of the
$\Bh=0$ solution.  We now see that this speculation is incorrect -- only magnetic fields
$\Bh \geq \Bh_c$ accomplish this.  
The extremal entropy turns on continuously below the critical magnetic field.  A curve that passes
through all the points to better than  $.5 \%$ accuracy is given by
\bea
\sh = {1\over 4\sqrt{6} } {\sqrt{1-\left({\Bh \over \Bh_c}\right)^2} \over (1+\Bh^2)^{7/4} }
\label{fit}
\eea
The prefactor of $1/(4\sqrt{6})$ is fixed by the value for the purely electrically charged black brane, 
which is of course known analytically. It is remarkable that this  function fits the data so well, 
but we are unable to say whether it represents an exact analytical result.  

\sm

To excellent 
accuracy, and as exhibited by (\ref{fit}),  we find 
\bea
\sh \sim \sqrt{\Bh_c - \Bh} 
\eea
near the critical point.     
If we increase the temperature at fixed magnetic field in this regime, we again find
a region obeying the same $\sh \sim \Th^{1/3}$ scaling as described in section 3.3.  

\subsection{Scaling region near the critical point}

The fact that the entropy density at the critical magnetic field has a simple power law dependence on 
temperature is indicative of a quantum critical point.  Recall that since we are working in terms
of dimensionless quantities as defined by the scalings in the asymptotically AdS$_5$ region,
{\sl a priori} the dimensionless entropy density  $\sh$ is allowed to be an arbitrary function of the dimensionless
temperature $\Th$.   This interpretation can be sharpened further by writing down a scaling form
for the entropy in the vicinity of the critical point
\bea
\label{fwithhat}
\sh = \Th^{1\over 3} f\left({\Bh-\Bh_c \over \Th^{2/3}}\right)
\eea
where, according to our results, the scaling function $f(x)$ has asymptotic behavior
\bea
f(x) \sim \left\{   \begin{array}{ccc} c_1 \sqrt{-x} && x\rightarrow - \infty \\
c_2 && x\rightarrow 0 \\
{c_3/ x} && x\rightarrow \infty \end{array}\right.
\label{scalingf}
\eea
for some constants $c_{1,2,3}$.  
In general, the entropy is a function of two dimensionless combinations of $B$, $\rho$, and 
$T$, whereas near the critical point the claim is that it can be written as a function of only one
variable.  All of this is of course standard from the general theory of classical and quantum 
critical phenomena.   

\sm

It is further useful to recall that for a quantum critical theory in $d$ spatial dimensions, 
with dynamical critical exponent $z$, and with a relevant coupling $m$ of scale dimension 
$\Delta$, the entropy density will take the scaling form 
\bea
s = T^{d/ z} f\left( {m^{2/ \Delta} \over T^{2/z}}\right)
\eea
Comparing to (\ref{scalingf}) we read off $d=1$, $z=3$, and $\Delta =2$. 

\sm

\begin{figure}[h!]
\begin{centering} 
\includegraphics[scale=.6]{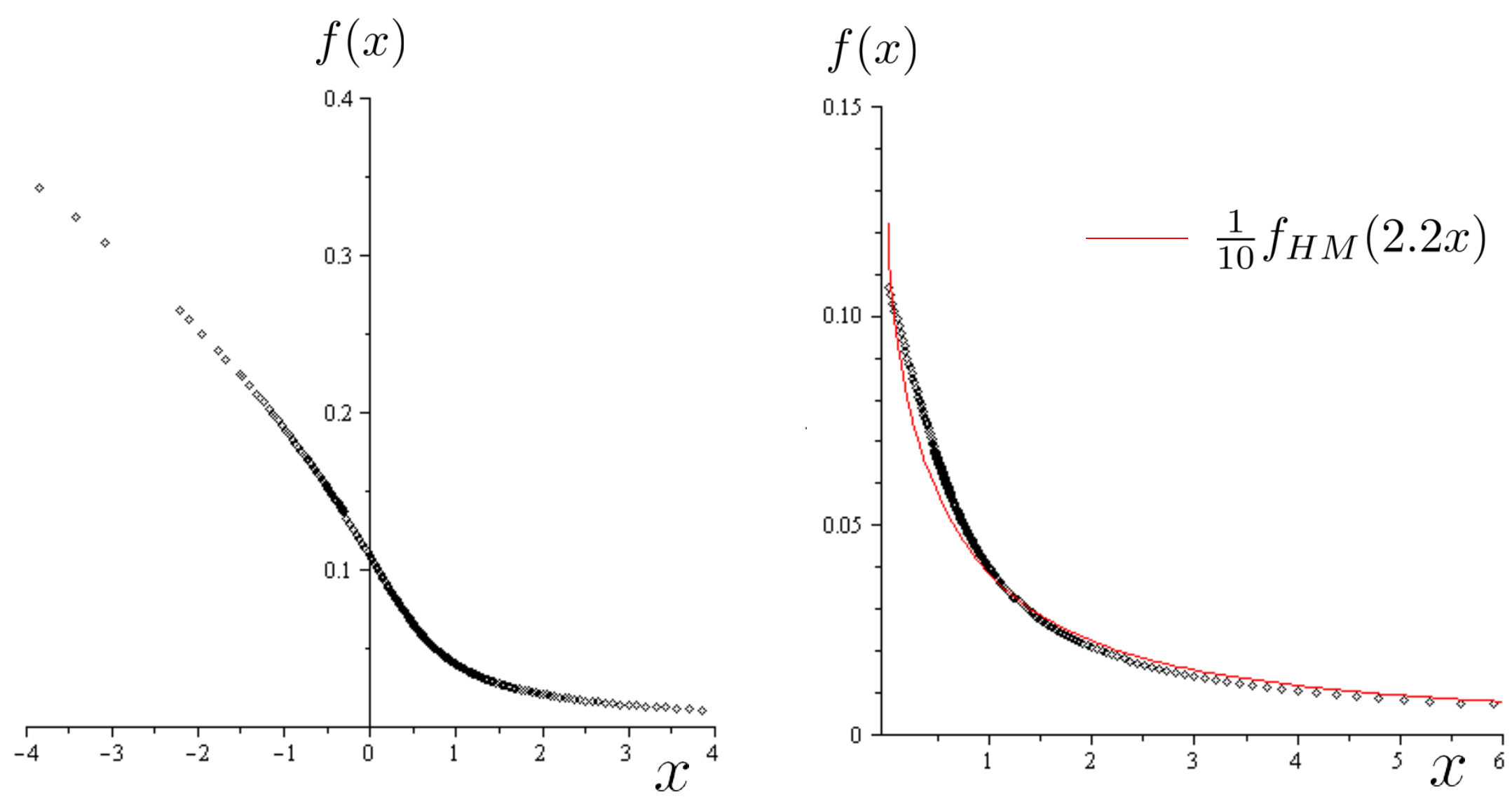}
\caption{Plots of the  scaling function $f(x)$ controlling the thermodynamics 
within the scaling region.  On the right plot we restrict to $x>0$, corresponding 
to $\Bh>\Bh_c$, and compare the scaling function from gravity with a free field 
computation based on the Hertz/Millis theory; the latter is displayed as a red line.   
The factors of $1/10$ and $2.2$  in the latter are chosen to give 
a good match, and are expected since the the gravitational result contains a factor of 
Newton's constant and the normalizations of the magnetic fields need not agree.  }
\label{scaling}
\end{centering}
\end{figure}

We have verified that the scaling form (\ref{scalingf}) conforms to  our numerical
results  near the critical point at $\hat B=\hat B_c$ and $ \hat T=0$.  
From the numerics we can reconstruct the form of 
the scaling function, which is shown in Figure \ref{scaling}.
We can use this to determine the constants appearing in  (\ref{scalingf}) to be,
\bea
c_1 \approx 0.172 \hskip 1in  c_2 \approx 0.110 \hskip 1in c_3\approx 0.045
\eea 

As noted in the Introduction, the standard approach to modelling a magnetically tuned
quantum critical point of the type that we are seeing is based on the Hertz/Millis theory.
We consider the action in (\ref{HM}) with $d=1$.    To the extent that this action captures the low
energy degrees of freedom of our theory, it is natural to compute its finite temperature 
entropy density  and compare to our results.  This cannot be entirely correct for a number 
of reasons, not least that  there is no way of explaining
the ground state entropy density in this framework, but the comparison is instructive nonetheless.
In the free field limit we just need the dispersion relation implied by (\ref{HM}), which is
\bea
\omega = |k|(k^2 +m^2) \hskip 1.2in m^2 = \Bh-\Bh_c
\eea
The partition function is 
\bea  \ln Z = - {L\over 2\pi} \int_{-\infty} ^\infty\! dk \ln(1-e^{-\beta |k| (k^2+m^2)})
\eea
from which we can extract the entropy density  as 
\bea
s  ={1\over L}  \left(1-\beta {\partial \over \partial \beta}\right)\ln Z 
\eea
We can write the result in a scaling form as
\bea
s = T^{1/3} f_{HM}\left( { \Bh-\Bh_c \over T^{2/3}}\right)
\eea
and then compare with the scaling function coming from gravity. The comparison is shown
in the right panel of Figure \ref{scaling}.   We only compare in the region $\Bh >\Bh_c$, since
it is only in this region that the Hertz/Millis action applies.   Nothing fixes the normalization of the
magnetic field appearing in  (\ref{HM}) compared to that in gravity, and so we have allowed
ourselves to adjust this relationship to achieve a good fit.  Similarly, we have introduced a
parameter to adjust the overall normalizations.   The functions match surprisingly  well, although it is
unclear how much significance to attach to this.

\subsection{The quantum critical point}

Initial data  yielding our closest approach to the quantum  critical point is given by
\bea
b& = & 1.7320507
\no \\
q & = & 5.08\times 10^{-4}
\label{bq}
\eea
This choice of $(b,q)$ along with the gauge choice $C'(r_+)=3.464063$  leads to 
the following values for the asymptotic parameters of (\ref{asymdata}),
\bea
v&=& 5.23811\times 10^{-14},
\hskip 0.3in  
w= 1.60044\times 10^{-9}
\hskip 0.3in c_0= 859.258
\no \\
e_3&=&  5.20520\times10^{20}
\hskip 0.43in  p_3=  -2.08236\times 10^{16}
\label{params}
\eea
from which we can compute the temperature and entropy density as 
\bea
\Th & = & 9.59602\times 10^{-9}
\no \\
\sh & = & 2.08699\times 10^{-4}
\eea
From (\ref{bq}) we see that there is a large hierarchy between the size of
the electric and magnetic charges at the horizon in our chosen coordinates.   Nevertheless,
measured at infinity in the rest frame the ratio of these quantities is of order 
unity, $\hat B_c^3 \approx 0.124569$.  This happens because there are large rescalings 
that occur in the process of integrating out from the horizon to the asymptotic regions, as 
illustrated by the magnitude of the parameters in  (\ref{params}).    Using  (\ref{phys}) these can convert a
large hierarchy into one of order unity.   

\begin{figure}[h!]
\begin{centering} 
\includegraphics[scale=.6]{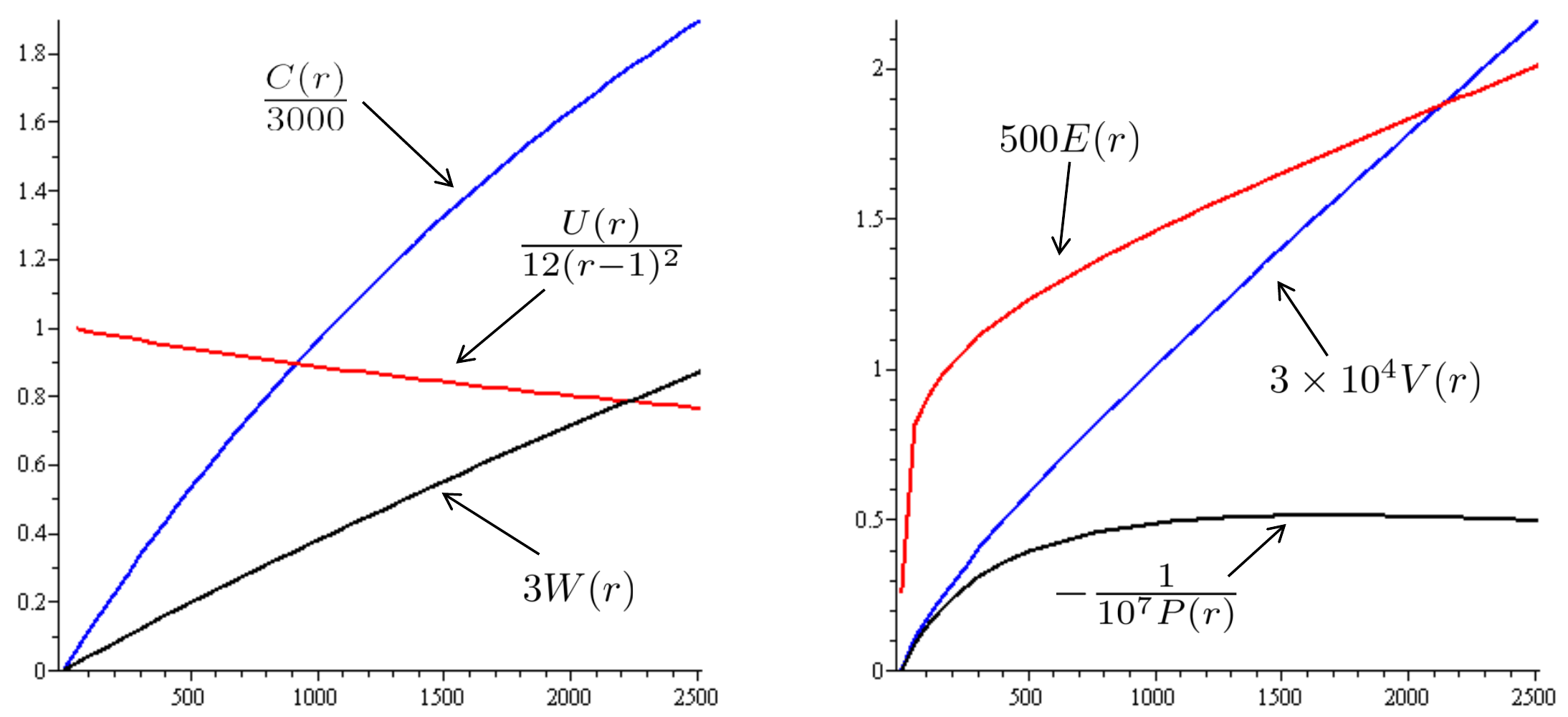}
\caption{Plots of the metric functions for a solution very near the quantum critical point.  
Multiplicative factors have been included to allow the functions to appear on the same plot. 
The form of the functions near the horizon show that the metric is behaving as AdS$_3 \times \bR^2$
plus small corrections.   These small corrections are amplified in passing to the asymptotic 
AdS$_5$ regime.  }
\label{metric}
\end{centering}
\end{figure}

It is instructive to examine the metric functions for this near critical solution.  
In Figure~\ref{metric}
we provide plots of these functions in the near (and not so near) horizon region.  
Near the horizon, one can think of these metric functions  as representing a deformation of the 
configuration
\bea
U(r) & = & 12(r-1)^2 \hskip 1.05in  B=\sqrt{3}
\no \\
C(r)& = &  2\sqrt{3}(r-1) \hskip 1in  V=W=E=P=0
\label{AdS}
\eea
This represents AdS$_3 \times \bf R^2$ in ``boosted coordinates" with magnetic flux.  
We expect that, as we 
get nearer and nearer to the critical point, the metric functions will approach (\ref{AdS}) to
greater and greater accuracy.  On the other hand, it is crucial that there always be some
deviation away from (\ref{AdS}), as this solution is purely magnetic with vanishing electric
charge density.  Presumably what happens is that there is a fine-tuned limit in which all
of the charge is carried by the bulk supergravity fields outside the near horizon region; this 
is possible by virtue of the Chern-Simons term.   It would of course be extremely useful if
this limiting solution could be found analytically. 

\sm

We close this subsection with an issue which has not yet been conclusively resolved by 
our numerical studies. For large $\hat B$, all flows towards lower $\hat T$ end in
the purely magnetic fixed point solution of \cite{D'Hoker:2009mm}, whose near-horizon 
geometry is  $AdS_3 \times \bR^2$. More precisely, it can be shown numerically that
{\sl all flows for $\hat B > \hat B_c$ end up at the purely magnetic fixed point}. 
Where do the flows for $\hat B < \hat B_c$ end? Numerically, we have established 
that flows for $\hat B$ less than but close to $\hat B_c$ come near the purely magnetic 
fixed point, but  at temperatures so low (on the order of $\hat T \sim 10^{-15}$) that further numerical 
analysis can be pursued only at the cost of calculation times and memory which exceed
those of our computers. Thus, the logical possibility that the flows with $\hat B < \hat B_c$
end up on the critical curve, before the purely magnetic fixed point has been 
attained, cannot be excluded at this time. If this situation is in fact the one that occurs,
then it would have a surprising similarity to the dynamics of the $k=1$ theory,
derived in \cite{D'Hoker:2009bc}.

\subsection{Metamagnetic quantum criticality in  Sr$_3$Ru$_2$O$_7$  } 

We close this section with a few words about a real system that parallels ours in 
a number of ways.  Quantum criticality and novel phases  in  Sr$_3$Ru$_2$O$_7$ 
have been the subject of much experimental and theoretical interest in the past few years; e.g.     
\cite{RPMMG,FKLEM}. 
Sr$_3$Ru$_2$O$_7$ is a layered structure, which for a large magnetic field 
perpendicular to the layers exhibits a line of first order metamagnetic phase 
transitions at finite temperature, ending at a finite temperature critical point.  By 
including a component of magnetic field in the plane of the layers, the critical 
point can be brought to zero temperature.  As in our case, the transition occurs 
at finite magnetic field,  and involves no change of symmetry.   Away from the 
critical point the system behaves as a Fermi liquid,
with entropy density linear in temperature.  As the critical point is approached, the 
linear term diverges as $s/T \sim 1/(B-B_c)$, just as  we found (in this case $B_c$ is 
approximately $8$ Tesla.)    In recent work, the complete ``entropy landscape"
of Sr$_3$Ru$_2$O$_7$ at finite temperature and magnetic field has been mapped 
out \cite{RPMMG}.  In very pure samples, as one tries to sit right on top of the 
critical point one finds instead that a new phase emerges,
which is believed to be a spatially anisotropic nematic phase \cite{OKE,FKLEM}.   
This has been described as nature's solution
to the problem of avoiding a finite entropy density at zero temperature.   
The parallels with our system are evident (though an obvious
difference is that our critical theory effectively sees only a single spatial direction, compared to the
two in-plane directions in Sr$_3$Ru$_2$O$_7$), and lead us to speculate whether in our case
 the extremal black hole phase is unstable and gives way to a new phase with reduced symmetry, as
in \cite{Nakamura:2009tf}. 
 
\section{Discussion}

We have found a holographic description of a quantum critical point, reached by tuning
a magnetic field at finite density,  that  nicely resembles examples seen in the real world.  
In particular, approached from the high $B$-field side, we have Fermi liquid behavior; 
as the magnetic field is lowered the
specific heat coefficient diverges, and we enter a regime of non-Fermi liquid behavior with
nontrivial scaling properties. 

\sm

 The most exotic property as compared to known physical examples
is the existence of a ground state entropy on the low field side of the transition.
The presence of a non-vanishing entropy at $T=0$ for $\hat B < \hat B_c$ would again 
seem to contradict the third ``law" of thermodynamics \cite{D'Hoker:2009bc},
as it did in the absence of magnetic fields. Actually, our results show that
the lifting of the ground state entropy as a function of an external magnetic field
is a subtle and dynamical issue. Also, we expect the ground state entropy 
to be lifted by instabilities driving the system towards inhomogeneity, or by turning on further fields besides an external magnetic field.

\sm

It is worth emphasizing again the universality of our results: they apply to {\sl all} 
supersymmetric AdS$_5$ examples arising from IIB or M-theory, since all such 
theories admit a consistent truncation to the  Einstein-Maxwell-Chern-Simons theory used
here \cite{Buchel:2006gb,Gauntlett:2006ai,Gauntlett:2007ma}.   We did not have to introduce 
any model building devices in the way of probe branes
or scalar fields.  

\sm

There are many open questions and avenues for further investigation.  Many of our numerical results
cry out for an analytical derivation.   In particular, one would expect to be able to derive the
value $z=3$ of the dynamical exponent, and the dimension $\Delta =2$ of the relevant operator
governing the critical theory.   It may similarly be possible to understand these results microscopically 
on the gauge theory side. 

\sm

All of our results were presented for the supersymmetric value of the Chern-Simons coupling,
$k=2/\sqrt{3}$, but it would be useful to consider other values as well.  Our expectation is that
as $k$ is increased the critical point will retain its character but with  $\Bh_c$
moving to a smaller value.  An interesting question would then  be whether we reach $\Bh_c=0$ 
at finite $k$, for if so there would no longer be an extremal black hole phase. 

\sm

In this work we have only considered the thermodynamics, but the existence of the critical
theory implies scaling behavior of correlation functions with respect to frequency and momentum. 
Computing these would be valuable in pinning down the precise connection to the Hertz/Millis theory.

\sm

Finally, to more closely model real systems it would be very interesting to construct a 
version of our system giving rise to critical behavior in two or three spatial dimensions. 

\bigskip \bigskip 

\noindent
{\Large \bf Acknowledgments}

\medskip

It is a pleasure to thank David Berenstein, Sudip Chakravarty, Tom Faulkner, Gary Horowitz, Don Marolf,  
Eric Perlmutter, Joe Polchinski,  and  Matt Roberts  for helpful discussions.


\begin{thebibliography}{99} 

\bibitem{Buchel:2006gb}
  A.~Buchel and J.~T.~Liu,
  ``Gauged supergravity from type IIB string theory on Y(p,q) manifolds,''
  Nucl.\ Phys.\  B {\bf 771} (2007) 93
  [arXiv:hep-th/0608002].

\bibitem{Gauntlett:2006ai}
  J.~P.~Gauntlett, E.~O Colgain and O.~Varela,
  ``Properties of some conformal field theories with M-theory duals,''
  JHEP {\bf 0702} (2007) 049
  [arXiv:hep-th/0611219].

\bibitem{Gauntlett:2007ma}
  J.~P.~Gauntlett and O.~Varela,
  ``Consistent Kaluza-Klein Reductions for General Supersymmetric AdS
  Solutions,''
  Phys.\ Rev.\  D {\bf 76} (2007) 126007
  [arXiv:0707.2315 [hep-th]].

\bibitem{sachdev}
S.~Sachdev, ``Quantum Phase Transitions",  Cambridge University Press (2001)

\bibitem{Liu:2009dm}
  H.~Liu, J.~McGreevy and D.~Vegh,
  ``Non-Fermi liquids from holography,''
  arXiv:0903.2477 [hep-th].

\bibitem{Cubrovic:2009ye}
  M.~Cubrovic, J.~Zaanen and K.~Schalm,
  ``Fermions and the AdS/CFT correspondence: quantum phase transitions and the
  emergent Fermi-liquid,''
  arXiv:0904.1993 [hep-th].

\bibitem{Faulkner:2009wj}
  T.~Faulkner, H.~Liu, J.~McGreevy and D.~Vegh,
  ``Emergent quantum criticality, Fermi surfaces, and AdS2,''
  arXiv:0907.2694 [hep-th].

\bibitem{Faulkner:2010tq}
  T.~Faulkner and J.~Polchinski,
  ``Semi-Holographic Fermi Liquids,''
  arXiv:1001.5049 [hep-th].

\bibitem{Rey:2008zz}
  S.~J.~Rey,
  ``String Theory On Thin Semiconductors: Holographic Realization Of Fermi
 Points And Surfaces,''
  Prog.\ Theor.\ Phys.\ Suppl.\  {\bf 177}, 128 (2009)
  [arXiv:0911.5295 [hep-th]].


\bibitem{D'Hoker:2009mm}
  E.~D'Hoker and P.~Kraus,
  ``Magnetic Brane Solutions in AdS,''
  JHEP {\bf 0910}, 088 (2009)
  [arXiv:0908.3875 [hep-th]].

\bibitem{D'Hoker:2009bc}
  E.~D'Hoker and P.~Kraus,
  ``Charged Magnetic Brane Solutions in $AdS_5$ and the fate of the third law of
  thermodynamics,''
  arXiv:0911.4518 [hep-th].

\bibitem{Anninos:2008fx}
  D.~Anninos, W.~Li, M.~Padi, W.~Song and A.~Strominger,
  ``Warped AdS$_3$ Black Holes,''
  JHEP {\bf 0903}, 130 (2009)
  [arXiv:0807.3040 [hep-th]].

\bibitem{Compere:2009zj}
  G.~Compere and S.~Detournay,
  ``Boundary conditions for spacelike and timelike warped AdS$_3$ spaces in
  topologically massive gravity,''
  JHEP {\bf 0908}, 092 (2009)
  [arXiv:0906.1243 [hep-th]].

\bibitem{Millis02}
A. J. Millis, A. J. Schofield, G. G. Lonzarich and S. A. Grigera, ``Metamagnetic quantum
criticality in metals," Phys. Rev. Lett. {\bf 88}, 217204 (2002).

\bibitem{RPMMG}
A. W. Rost, R. S. Perry, J.-F. Mercure, A. P. Mackenzie, and  S. A. Grigera
``Entropy Landscape of Phase Formation Associated with Quantum Criticality in $Sr_3Ru_2O_7$",
 Science
Vol. 325. no. 5946, pp. 1360 - 1363 (2009)

\bibitem{Lippert}
  G.~Lifschytz and M.~Lippert,
  ``Holographic Magnetic Phase Transition,''
  Phys.\ Rev.\  D {\bf 80}, 066007 (2009)
  [arXiv:0906.3892 [hep-th]].


\bibitem{hertz}
J. A. Hertz, ``Quantum critical phenomena," Phys. Rev. {\bf B 14}  1165 (1976).

\bibitem{millis93}
A. J. Millis, ``Effect of a nonzero temperature on quantum critical points in itinerant fermion systems",
Phys. Rev. {\bf  B 48}, 7183–7196 (1993) 

\bibitem{LRMW}
H. v. Lohneysen, A. Rosch, M. Mojta, and P. Wolfle, ``Fermi-liquid instabilities at magnetic 
quantum phase transitions", 
Rev. Mod. Phys. {\bf 79} , 1015–1075  (2007) 

\bibitem{Witten:1998zw}
  E.~Witten,
  ``Anti-de Sitter space, thermal phase transition, and confinement in  gauge
  theories,''
  Adv.\ Theor.\ Math.\ Phys.\  {\bf 2}, 505 (1998)
  [arXiv:hep-th/9803131].


\bibitem{Chamblin:1999tk}
  A.~Chamblin, R.~Emparan, C.~V.~Johnson and R.~C.~Myers,
  ``Charged AdS black holes and catastrophic holography,''
  Phys.\ Rev.\  D {\bf 60}, 064018 (1999)
  [arXiv:hep-th/9902170].



\bibitem{Parnachev:2006dn}
  A.~Parnachev and D.~A.~Sahakyan,
  ``Chiral phase transition from string theory,''
  Phys.\ Rev.\ Lett.\  {\bf 97}, 111601 (2006)
  [arXiv:hep-th/0604173].

\bibitem{Mateos:2006nu}
  D.~Mateos, R.~C.~Myers and R.~M.~Thomson,
  ``Holographic phase transitions with fundamental matter,''
  Phys.\ Rev.\ Lett.\  {\bf 97}, 091601 (2006)
  [arXiv:hep-th/0605046].



\bibitem{Davis:2007ka}
  J.~L.~Davis, M.~Gutperle, P.~Kraus and I.~Sachs,
  ``Stringy NJL and Gross-Neveu models at finite density and temperature,''
  JHEP {\bf 0710}, 049 (2007)
  [arXiv:0708.0589 [hep-th]].

\bibitem{Hartnoll:2008vx}
  S.~A.~Hartnoll, C.~P.~Herzog and G.~T.~Horowitz,
  ``Building a Holographic Superconductor,''
  Phys.\ Rev.\ Lett.\  {\bf 101}, 031601 (2008)
  [arXiv:0803.3295 [hep-th]].                    

\bibitem{Davis:2008nv}
  J.~L.~Davis, P.~Kraus and A.~Shah,
  ``Gravity Dual of a Quantum Hall Plateau Transition,''
  JHEP {\bf 0811}, 020 (2008)
  [arXiv:0809.1876 [hep-th]].




\bibitem{Evans:2010iy}
  N.~Evans, A.~Gebauer, K.~Y.~Kim and M.~Magou,
  ``Holographic Description of the Phase Diagram of a Chiral Symmetry Breaking
  Gauge Theory,''
  arXiv:1002.1885 [hep-th].


\bibitem{Jensen:2010vd}
  K.~Jensen, A.~Karch and E.~G.~Thompson,
  ``A Holographic Quantum Critical Point at Finite Magnetic Field and Finite
  Density,''
  arXiv:1002.2447 [hep-th].

\bibitem{Horowitz:2009ij}
  G.~T.~Horowitz and M.~M.~Roberts,
  ``Zero Temperature Limit of Holographic Superconductors,''
  JHEP {\bf 0911}, 015 (2009)
  [arXiv:0908.3677 [hep-th]].


\bibitem{Goldstein:2009cv}
  K.~Goldstein, S.~Kachru, S.~Prakash and S.~P.~Trivedi,
  ``Holography of Charged Dilaton Black Holes,''
  arXiv:0911.3586 [hep-th].

\bibitem{Cadoni:2009xm}
  M.~Cadoni, G.~D'Appollonio and P.~Pani,
  ``Phase transitions between Reissner-Nordstrom and dilatonic black holes in
  4D AdS spacetime,''
  JHEP {\bf 1003}, 100 (2010)
  [arXiv:0912.3520 [hep-th]].


\bibitem{Gauntlett:2009bh}
  J.~Gauntlett, J.~Sonner and T.~Wiseman,
  ``Quantum Criticality and Holographic Superconductors in M-theory,''
  JHEP {\bf 1002}, 060 (2010)
  [arXiv:0912.0512 [hep-th]].



\bibitem{OKE}
V. Oganesyan, S. A. Kivelson and E. Fradkin, ``Quantum theory of a nematic Fermi
fuid," Phys. Rev. {\bf B 64}, 195109 (2001).

\bibitem{FKLEM}
 E. Fradkin, S.A. Kivelson, M. J. Lawler, J. P. Eisenstein, and A. P. Mackenzie, 
 ``Nematic Fermi Fluids in Condensed Matter Physics", arXiv:0910.4166

\bibitem{Nakamura:2009tf}
  S.~Nakamura, H.~Ooguri and C.~S.~Park,
  ``Gravity Dual of Spatially Modulated Phase,''
  arXiv:0911.0679 [hep-th].


 
\end{thebibliography}
\end{document}